\documentclass[conference]{IEEEtran}
\IEEEoverridecommandlockouts
\usepackage{cite}
\usepackage{amsmath,amssymb,amsfonts}
\usepackage{algorithmic}
\usepackage{graphicx}
\usepackage{textcomp}
\usepackage{xcolor}

\usepackage{caption}
\usepackage{subcaption}

\def\BibTeX{{\rm B\kern-.05em{\sc i\kern-.025em b}\kern-.08em
    T\kern-.1667em\lower.7ex\hbox{E}\kern-.125emX}}
\begin{document}

\title{Doppler Ambiguity Elimination Using 5G Signals in Integrated Sensing and Communication\vspace*{-1mm} \\
\thanks{This work was funded by the European Commission through the H2020 MSCA 5GSmartFact project under grant agreement number 956670.}
}

\author{Keivan Khosroshahi\IEEEauthorrefmark{1}\IEEEauthorrefmark{2}, Philippe Sehier\IEEEauthorrefmark{2}, and Sami Mekki\IEEEauthorrefmark{2},\\
\IEEEauthorrefmark{1}
Université Paris-Saclay, CNRS, CentraleSupélec, Laboratoire des Signaux et Systèmes, Gif-sur-Yvette, France\\
\IEEEauthorrefmark{2} Nokia Standards, Massy, France
\\ keivan.khosroshahi@centralesupelec.fr, \{philippe.sehier, sami.mekki\}@nokia.com
}

\maketitle
 
\begin{abstract}
The industrial point of view towards integrated sensing and communication (ISAC), the preference is to leverage existing resources and fifth-generation (5G) infrastructure to minimize deployment costs and complexity. In this context, we explore the utilization of current 5G new radio (NR) signals aligned with 3rd generation partnership project (3GPP) standards. Positioning reference signals (PRS) for sensing and physical downlink shared channel (PDSCH) for communication have been chosen to form an ISAC framework. However, PRS-based sensing suffers from Doppler ambiguity when the Doppler frequency shift is severe. To address this challenge, we introduce a novel method within the ISAC system that leverages the demodulation reference signal (DMRS) present in PDSCH to eliminate Doppler ambiguity. Furthermore, we formulate a resource allocation problem between PRS and PDSCH to achieve a Pareto optimal point between communication and sensing without Doppler ambiguity. Through simulations and analysis, we demonstrate the effectiveness of our proposed method on joint DMRS-PRS exploitation in mitigating Doppler ambiguity and the efficiency of the resource allocation scheme in achieving Pareto optimality for ISAC within a 5G NR framework.
\end{abstract}

\section{Introduction}
The adoption of integrated sensing and communication (ISAC) technology in upcoming mobile networks is increasingly gaining consensus due to the emergence of future applications in the domains of industry 4.0, automotive, internet of things (IoT) in smart cities and health care, which impose new requirements on wireless communication networks \cite{zhang2021enabling}.
Integrating sensing services into the mobile infrastructure enables new services to be deployed at low cost, reusing the mobile coverage base already deployed and without requiring additional spectrum resources, while minimizing the impact on existing equipment.

Several reference signals have been defined in fifth-generation (5G) new radio (NR), addressing different objectives. They have been specified to provide excellent passive detection performance, robust anti-noise capabilities, and favorable auto-correlation characteristics \cite{wei20225g}. Moreover, using pilots in ISAC systems leads to low hardware costs, easy implementation, and low computational complexity \cite{ozkaptan2018ofdm}. Consequently, significant interest has emerged in designing ISAC signals based on pilots. Among the pilots available in 5G networks, the use of the positioning reference signal (PRS) for sensing purposes holds significant promise, given its rich time-frequency resources and flexible configuration. PRS, introduced in 3rd generation partnership project (3GPP) Release 16 of the 5G specification, aims to improve the positioning accuracy of connected user equipments (UEs) due to its high resource element (RE) density and superior correlation properties compared to existing reference signals due to the diagonal or staggered PRS RE pattern.  In sensing context, PRS is used in \cite{wei20225g} to estimate the range and Doppler, and its performance is compared with other pilots available in 5G standard, namely demodulation reference signal (DMRS), channel state information reference signal (CSI-RS) and synchronization signal (SS). A two-stage joint pilot optimization, target detection, and channel estimation scheme has been introduced in \cite{huang2022joint} for ISAC. In \cite{ma2022downlink}, DMRS and CSI-RS are separately used for velocity and range estimation. In \cite{wang2020multi}, a multi-range joint automotive radar and communication system based on pilot-based orthogonal frequency-division multiplexing (OFDM) waveform is investigated. PRS and DMRS are jointly used in \cite{khosroshahi2024leveragingprspdschintegrated} to remove range ambiguity in ISAC scenarios.

However, exploiting the existing reference signals of the 5G-NR raises new issues. This is because the sensed targets might be far from the gNB, or their speed may be high. Consequently, the reflected signals may be received with a large time delay, above the cyclic prefix length, and large Doppler shifts. 
Periodogram-based estimation for delay and Doppler exhibit spurious peaks in these situations, creating ambiguities in distance or velocity estimates. Such fake peaks can be interpreted as real targets. While the proposed solution in \cite{wei2023multiple} has focused on multiple reference signals to address the ambiguity, they are not directly applicable in ISAC scenarios due to potential interference between physical downlink shared channel (PDSCH) for data transmission and other reference signals such as PRS and CSI-RS. Additionally, the method described in \cite{wei20225g} leads to a significant reduction in the maximum detection range, especially when employing large PRS comb sizes. Hence, to the best of our knowledge, a 3GPP-compliant solution to address range ambiguity within the ISAC framework has not been proposed.

In this work, we introduce a novel approach leveraging the current reference signals in 5G NR to enable sensing for unconnected targets with enhanced maximum detectable Doppler frequency shift without Doppler ambiguity. To provide sensing services alongside data transmission, we propose an ISAC OFDM resource grid comprising PDSCH for communication and PRS for sensing. To remove the Doppler ambiguity, we propose a method to reuse DMRS within PDSCH for sensing in addition to communication. Furthermore, we propose time and frequency resource allocation between PRS and PDSCH since sensing and communication symbols cannot share the same REs in an OFDM resource grid. Building upon this, we formulate a resource allocation problem between PDSCH and PRS to achieve Pareto optimality between communication and sensing without Doppler ambiguity. Ultimately, via extensive simulations, we prove the efficacy of our proposed method in addressing Doppler ambiguity and finding the Pareto optimal point between communication and sensing.

\section{PRS AND DMRS DESCRIPTION}
The 5G-NR waveform is built on a time-frequency resource grid as described in technical specifications (TS) 38.214 \cite{3gpp2018nr1} and the smallest time-frequency resource is a physical resource
block (PRB), consisting of 12 contiguous sub-carriers and 14 consecutive symbols. In the following two sections, we explain about PRS and DMRS in 5G NR.

\subsection{PRS}
\begin{figure}[!t]
    \centering
    \begin{subfigure}{0.45\columnwidth}
        \centering
        \includegraphics[width=\linewidth]{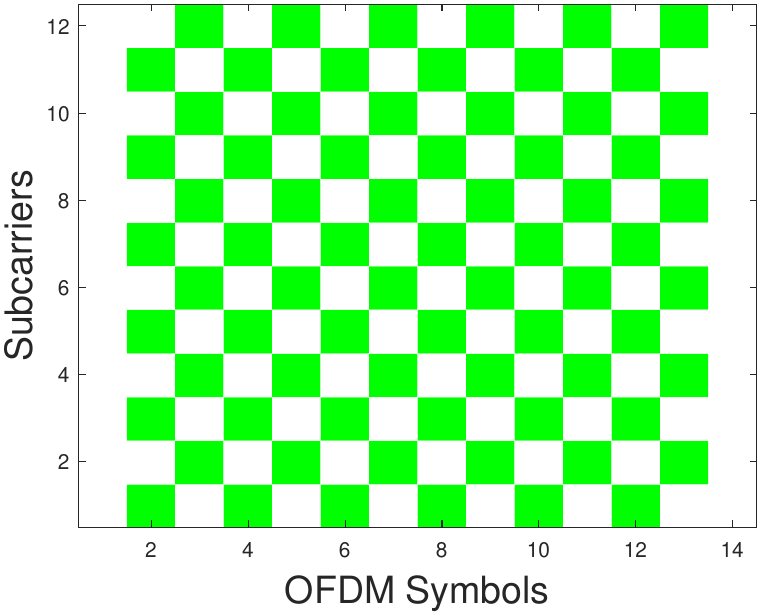}
        \caption{Comb size = 2}
        \label{comb4}
    \end{subfigure}
    \begin{subfigure}{0.45\columnwidth}
        \centering
        \includegraphics[width=\linewidth]{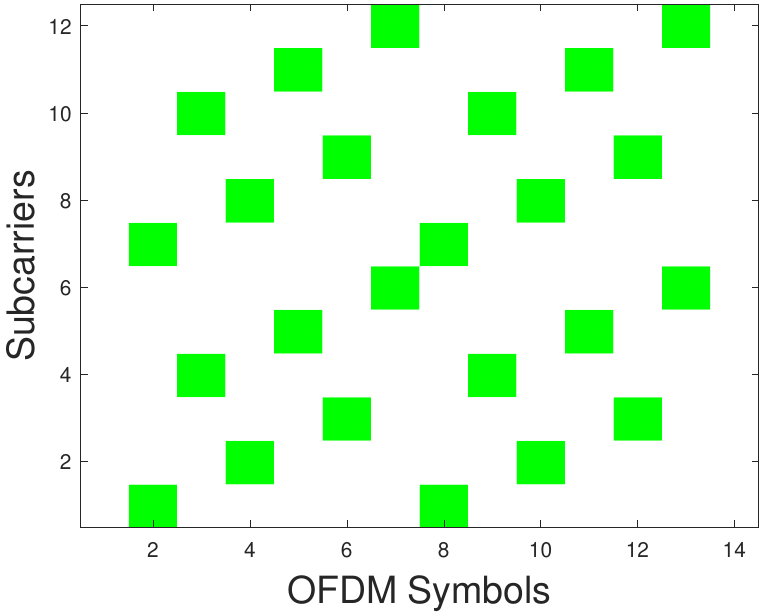}
        \caption{Comb size = 6}
        \label{comb6}
    \end{subfigure}
    \caption{PRS allocation with starting symbol= 2 and different Comb sizes.}
    \label{comb}
\end{figure}

The generation of PRS as explained in TS 38.211 \cite{3gpp2018nr} as follows
\vspace*{-1mm}
\begin{equation}
   r(m) = \frac{1}{\sqrt{2}}(1 - 2c(2m)) + j\frac{1}{\sqrt{2}}(1 - 2c(2m + 1)),
   \label{r}
\end{equation}

where $c(i)$ is denoted as Gold sequence of length-$31$ and the starting value of $c(i)$ for the PRS is provided in \cite{3gpp2018nr}. PRS is generated from the Gold sequence and it has favorable anti-noise and auto-correlation characteristics. The PRS is allocated a minimum of 24 PRBs and a maximum of 272 PRBs, depending on the carrier size, showing the flexible transmission parameters supported in 5G NR. This flexibility allows for a versatile configuration of time-frequency resources to fulfill sensing accuracy requirements across diverse applications and scenarios. As outlined in TS 38.211 \cite{3gpp2018nr}, PRS supports four comb structures, i.e., Comb $\{2,4,6,12\}$, in the frequency domain and five symbol number configurations, i.e., Symbol $\{1,2,4,6,12\}$, in the time domain. The structure of PRS with 12 symbols and two comb sizes is illustrated in Fig. \ref{comb}.

\subsection{DMRS}
Another reference signal 
that can potentially be used in the ISAC framework is DMRS, available in the PDSCH. DMRS replaces the cell-specific reference signal (CRS) in long-term evolution (LTE) for channel estimation, and PDSCH decoding \cite{3gpp2018nr}. DMRS generation is also based on \eqref{r}, and the initial value of $c(i)$ for DMRS can be found in \cite{3gpp2018nr}. DMRS generation occurs within the designated PDSCH allocation, as detailed in TS 38.211 \cite{3gpp2018nr}. The allocated resources for PDSCH are available in the bandwidth part (BWP) of the carrier as explained in TS 38.214 \cite{3gpp2018nr1}. DMRS is used for channel estimation and is confined to the resource blocks (RBs) assigned for PDSCH. Its structure is adaptable to various deployment scenarios and use cases. The positioning of DMRS symbols depends on the mapping type, which can be slot-wise (Type A) or non-slot-wise (Type B). Additional DMRS symbol positions are determined by tables outlined in TS 38.211 \cite{3gpp2018nr}. Furthermore, front-load DMRS may occupy 1 to 4 OFDM symbols in the time domain. DMRS is within physical channels allocated for communication receivers and is transmitted using the same beam selected for communication receivers. 

\vspace*{-1mm}

\section{ISAC System Model}
\begin{figure}[!t]
\centering
\mbox{\includegraphics[width=\linewidth]{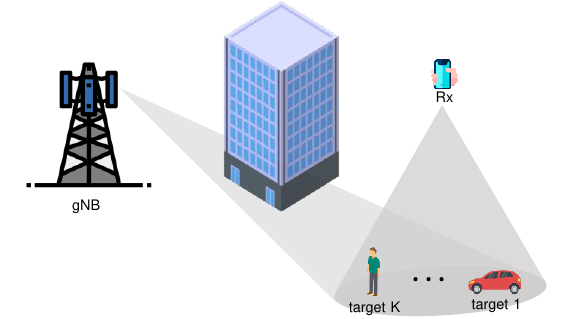}}
\caption{ISAC system model.}
\label{bistatic RIS2}
\end{figure}

The existing 5G NR signals were not initially designed for radar sensing applications. This section outlines a methodology for efficiently leveraging both communication and sensing signals within the ISAC framework. Our vision involves implementing a downlink ISAC system where a gNB directs signals toward a sensing area where $K$ point-like targets exist. The receiver captures echoes from these targets to estimate their range and velocity. We assume that the targets are within the line-of-sight (LOS) of both the transmitter and the receiver. Fig. \ref{bistatic RIS2} depicts an example of the considered scenario. If we consider an OFDM resource grid with $N$ symbols in the time domain and $M$ sub-carriers in the frequency domain, the transmitted signal in the time domain can be written in the following form: \cite{buzzi2019using}

\vspace*{-1mm}
\begin{equation}
    s(t) = \sum_{n = 0}^{N-1}\text{rect}(\frac{t - nT_0}{T_0})\sum_{m=0}^{M-1} v(m,n) e^{j2\pi m\Delta f(t - nT_0)},
\end{equation}
where $\text{rect}(t/T_0)$ denotes the rectangular pulse, $\Delta f$ represents the subcarrier spacing. $T_0 = T_{CP} + T_s$ is the total duration of the OFDM symbol, where $T_{s} = \frac{1}{\Delta f}$ represents the symbol duration, and $T_{CP}$ denotes the duration of the cyclic prefix (CP). $v(m,n)$ is the complex transmitted symbol at OFDM symbol $n$ and sub-carrier $M$, where $n=0,...,N-1$ and $m=0,...,M-1$ within an $M \times N$ OFDM resource grid. If there are $k$ targets present in the sensing area, the echo signals from targets received by the receiver can be written as \cite{braun2014ofdm}
\vspace*{-1mm}
\begin{align}
    r(t)=\sum_{k=1}^K\beta_k s(t-\tau_k)e^{j2\pi f_{d,k}t} + u(t),
    \label{y_time}
\end{align}
where $\beta_k$ denotes the attenuation factor of the $k$-th target, $f_{d,k}$ represents the Doppler frequency shift of the $k$-th target, $\tau_k$ signifies the delay of the $k$-th target and $u(t) \in \mathbb{C}$ refers to complex additive white Gaussian noise (AWGN) with zero mean and variance of $2\sigma^2$.

Conventional OFDM receivers proceed by sampling the signal at each symbol time, followed by performing fast fourier transform (FFT) to extract the modulated symbols. Under normal circumstances, delay and Doppler effects are compensated, leading to mitigation of intersymbol interference (ISI). The cyclic prefix is typically leveraged to address receive timing errors and multipath spread. However, in our scenario, the path delay and Doppler may not fall within the usual range due to significant uncertainties regarding the target's position and speed. As a result, it becomes necessary to consider ISI in both the time and frequency domains. Consequently, the received modulation symbols can be expressed as:
\vspace*{-1mm}
\begin{align}
    y(m,n) =&  \sum_{k=1}^K\beta_k e^{j2\pi n T_0 f_{d,k}} e^{-j2\pi m \Delta f \tau_k} v(m,n) + \text{ISI} \notag\\
    &+ q(m,n).
    \label{y_freq}
\end{align}

The term $q(m,n) \in \mathbb{C}$ represents the complex AWGN on the $m$-th subcarrier and the $n$-th OFDM symbol. This noise has a zero mean and a variance of $2\sigma^2$, and it is obtained from the sampling and FFT process over $r(t)$.

\section{Estimation of the range and velocity \label{est}}
Doppler and delay would cause a rotation on the time and frequency axis, respectively over the received samples. Determining range and Doppler necessitates a two-dimensional search. Multiple techniques are outlined in \cite{braun2014ofdm}; however, we focus here on a computationally efficient method operating in two stages: initially estimating the range and subsequently the Doppler, after compensating for the delay. The bistatic distance, i.e., from the transmitter to the target and then from the target to the receiver, can be derived from range assessment using the periodogram \cite{pucci2021performance}. To achieve this, we eliminate the transmitted sensing symbols from the received echoes through point-wise division as described below. 
\vspace*{-1mm}
\begin{align}
    g(m,n) =    
    \begin{cases}
      \frac{y(m,n)}{v(m,n)}, & \text{    If  } v(m,n) \neq 0,\\
      0, & \text{    If  } v(m,n) = 0.
    \end{cases}\label{gmn}
\end{align}

Afterward, we compute $M$-point inverse fast fourier transform (IFFT) of the PRS on the $n$-th column of the $g(m,n)$
\vspace*{-1mm}
\begin{align}
    r_n(l) =& |\text{IFFT}(g(m,n))| = |\sum_{k=1}^K (\beta_k e^{j2\pi n T_0 f_{d,k}} \sum_{m=0}^{M-1} \notag \\ 
    & e^{-j2\pi m \Delta f  \frac{R^{tot}_k}{c}} e^{j2\pi\frac{ml}{M}}+ \frac{q'(m,n)}{v(m,n)}e^{j2\pi\frac{ml}{M}})|,
    \label{r_l}
\end{align}
where $l= 0,...,M-1$, $|.|$ is the absolute value, $q'(m,n) = q(m,n) + \text{ISI}$, and we replaced $\tau_k = \frac{R^{tot}_k}{c}$ while $R^{tot}_k$ is the bistatic range, $c$ is the light speed. The absolute value is taken to reduce Doppler sensitivity. When the arguments of $e^{-j2\pi m \Delta f  \frac{R^{tot}_k}{c}} e^{j2\pi\frac{ml}{M}}$ in \eqref{r_l} cancel each other, the maximum value occurs. The next step is to perform IFFT over all columns of the $g(m,n)$ and average over them to increase the signal-to-noise ratio (SNR):
\vspace*{-1mm}
\begin{align}
   \overline{\rm r}(l) = \frac{1}{N}\sum_{n=0}^{N-1} r_n(l).
    \label{r_l2}
\end{align}

Afterward, we need to find the index of the maximum value named $\hat{l}_k$ in \eqref{r_l2}. Then, the bistatic distance of each target can be calculated as \cite{wei20225g} 
\vspace*{-1mm}
\begin{equation}
    \hat{R}^{tot}_k = \frac{\hat{l}_kc}{\Delta f M}.
    \label{r_tot}
\end{equation}

The range resolution is
\vspace*{-1mm}
\begin{equation}
    \Delta R = \frac{c}{\Delta f M},
    \label{R_res}
\end{equation}
and the expression for the maximum detection range can be written as
\vspace{-1mm}
\begin{equation}
    R_{max} =  \frac{c}{\Delta f}.
    \label{R_MAX}
\end{equation}

According to \cite{wei20225g}, enhancing accuracy performance is achievable by increasing the number of IFFT points $m_a$ as shown below
\vspace*{-1mm}
\begin{align}
    r_n(l) =& |\text{IFFT}(g(m,n))| = |\sum_{k=1}^K (\beta_k e^{j2\pi n T_0 f_{d,k}} \sum_{m=0}^{m_aM-1} \notag \\ 
    & e^{-j2\pi m \Delta f  \frac{R^{tot}_k}{c}} e^{j2\pi\frac{ml}{m_aM}} + \frac{q'(m,n)}{v(m,n)}e^{j2\pi\frac{ml}{m_aM}})|.
    \label{r_ma}
\end{align}

In this case, \eqref{r_tot} will change into
\begin{equation}
    \hat{R}^{tot}_k = \frac{\hat{l}_kc}{m_a\Delta f M}. 
\end{equation}

As highlighted in \cite{li2019ofdm}, increasing $m_a$ results in range accuracy enhancement, albeit at the expense of heightened computational complexity.

After compensating the delay caused by the targets, for Doppler estimation, we perform $N$-points FFT on the $m$-th row of the $g(m,n)$
\vspace*{-1mm}
\begin{align}
    v_m(d) =& |\text{FFT}(g(m,n))| = |\sum_{k=1}^K (\beta_k \sum_{n=0}^{N-1} e^{j2\pi n T_0 f_{d,k}} e^{-j2\pi\frac{nd}{N}}\notag \\ 
    & + \frac{q'(m,n)}{v(m,n)}e^{-j2\pi\frac{nd}{N}})|,
    \label{v_m}
\end{align}
where $d= 0,...,N-1$. We perform FFT across all rows of the $g(m,n)$ and then average the results as shown below:
\vspace*{-1mm}
\begin{align}
   \overline{\rm v}(d) = \frac{1}{M}\sum_{m=0}^{M-1} v_m(d).
    \label{v_m2}
\end{align}

Afterward, we determine the index of the maximum value, denoted as $\hat{d}_k$ for target $k$ in \eqref{v_m2}, and the Doppler frequency can be estimated as \cite{pucci2021performance} 
\vspace*{-1mm}
\begin{equation}
    \hat{f}_{d,k} = \frac{\hat{d}_k}{T_s N}.
\end{equation}

The Doppler frequency resolution is
\vspace*{-1mm}
\begin{equation}
    \Delta\hat{f} = \frac{1}{T_s N}.\,
    \label{doppler_res}
\end{equation}
and the maximum detectable Doppler frequency estimation is 
\vspace*{-1mm}
\begin{equation}
    \hat{f}_{max} = \frac{1}{T_s}.
    \label{v_max}
\end{equation}

We can improve the accuracy of Doppler frequency estimation by increasing the FFT point similar to range estimation.

\subsection{Doppler Ambiguity}
\begin{figure}[!t]
\centering
\mbox{\includegraphics[width=\linewidth]{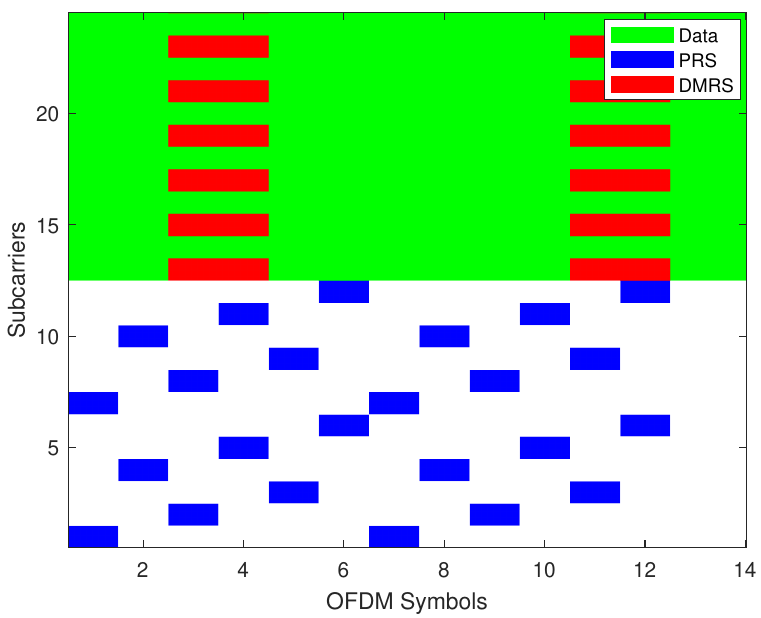}}
\caption{Resource allocation between PRS and PDSCH 
 showing a slot and 2 PRBs of the whole resource grid with $K_{comb} = 6$.}
\label{resourcegrid}
\end{figure}

PRS was initially designed for positioning in 5G NR and not for sensing. In positioning, time-frequency uncertainties are mitigated by the time-frequency servo loop, which is not the case for sensing. PRS features a periodic structure that can be adjusted by the comb size, as demonstrated in Fig. \ref{resourcegrid}, where an example with a comb size, i.e., $K_{comb} = 6$ is presented. 
The presence of empty REs within the PRS structure leads to the emergence of fake peaks in Doppler estimation using the method described in section \ref{est}. These peaks introduce ambiguity in distinguishing between the real and fake Doppler frequency shift resulting in Doppler ambiguity as shown in Figs. \ref{figa:PRS4} and \ref{figa:PRS12}. 
The cost of the approach proposed in \cite{wei20225g} to circumvent this ambiguity is the reduction of the maximum detectable range and Doppler frequency shift by a factor of $K_{comb}$. In scenarios where precise Doppler resolution is crucial, reducing the subcarrier spacing, as defined in \eqref{doppler_res}, becomes necessary. However, this adjustment with larger comb sizes leads to a notable reduction in the maximum detectable Doppler frequency shift. Hence, an alternative method is necessary to alleviate Doppler frequency ambiguity while ensuring a high maximum detectable Doppler frequency.

\subsection{Proposed Method}
The key idea to eliminate ambiguity in Doppler estimation is to leverage DMRS as a reference signal within the PDSCH for ISAC scenarios where the OFDM frame contains PRS for sensing and PDSCH for communication. To this end, we generate an OFDM frame, and based on the sensing and communication requirements of each application, we perform the time and frequency resource allocation. Fig. \ref{resourcegrid} depicts an example of such resource allocation in the frequency domain between PRS and PDSCH. In this scenario, DMRS serves not only for communication tasks like channel estimation but also for enhancing sensing capabilities by accurately distinguishing between real and fake Doppler frequency shifts. Additionally, this improvement is achieved without compromising accuracy or the maximum detectable Doppler frequency. Additionally, it does not require any modifications to the gNB configuration, thus fully aligning with 3GPP standards. 

Initially, we set double symbol for DMRS configuration and one additional DMRS symbol. Based on DMRS configuration type, the total number of subcarriers per PRB that can be allocated to DMRS, i.e., $M'$, is $4$ or $6$. We define $g_{DMRS}(m,n)$ and $g_{PRS}(m,n)$ as the result of point-wise division similar to \eqref{gmn} over DMRS and PRS symbols, respectively. Then, we form the following
\vspace{-3mm}
\begin{align}
    g_{PRS,DMRS}^{m,m'}(n) = g_{PRS}^{m}(n) + g_{DMRS}^{m'}(n),
\end{align}
where $g_{PRS}^{m}(n)$ is the $m$-th row of $g_{PRS}(m,n)$ and $g_{DMRS}^{m'}(n)$ is the $m'$-th row of $g_{DMRS}(m',n)$. Based on TS 38.211 \cite{3gpp2018nr}, $m' = \{0,2,4,6,8,10\}$ for DMRS configuration type $1$ and $m' = \{0,1,6,7\}$ for DMRS configuration type $2$. Then we calculate FFT over $m$-th row of $g_{PRS,DMRS}^{m,m'}(n)$ as follows

\vspace{-3mm}
\begin{align}
    v_m^{m'}&(d) = |\text{FFT}(g_{PRS,DMRS}^{m,m'}(n))| = |\sum_{k=1}^K (2\beta_k \sum_{n=0}^{N-1} e^{j2\pi n T_0 f_{d,k}}\notag \\
    &.e^{-j2\pi\frac{nd}{N}} + (\frac{q'(m,n)}{v_{PRS}(m,n)} + \frac{q'(m',n)}{v_{DMRS}(m',n)})e^{-j2\pi\frac{nd}{N}})|,
\end{align}
where $v_{PRS}$ and $v_{DMRS}$ are transmitted complex symbols of $PRS$ and $DMRS$, respectively. Afterward, we perform averaging over $m$ as follows
\vspace*{-1mm}
\begin{align}
   \overline{\rm v}^{m'}(d) = \frac{1}{M}\sum_{m=0}^{M-1} v_m^{m'}(d).
    \label{v_m3}
\end{align}
Eventually, we obtain the element wise multiplication of $\overline{\rm v}^{m'}(d)$ over $m'$ as follows
\begin{align}
   \overline{\rm v}(d) = \prod_{m'=0}^{M'-1} \overline{\rm v}^{m'}(d).
    \label{v_m4}
\end{align}

At the end, we use \eqref{v_m4} to find the index of the maximum value and estimate Doppler. The Doppler frequency shift resolution and the maximum detectable Doppler frequency can be obtained using \eqref{doppler_res} and \eqref{v_max}, respectively. This approach disrupts the periodic pattern of PRS, effectively reducing the magnitude of spurious peaks. Consequently, distinguishing between real and false Doppler frequencies becomes straightforward. 
If PRS and PDSCH have different resources allocated in the time domain, i.e., slots, we pad zeros to the reference signal with the smaller slots so $g_{PRS}^{m}(n)$ and $g_{DMRS}^{m'}(n)$ can have similar vector sizes.

\section{Pareto optimality in ISAC Resource Allocation}
In this section, we introduce an optimization problem to reach Pareto optimal point in time and frequency resource allocation between communication and sensing without Doppler ambiguity through the proposed algorithm. Towards this goal, we define the range resolution metric as $\frac{c}{\Delta f}$, $R_0$ as the maximum throughput of the received PDSCH per one slot and one PRB, and $\frac{1}{T_s}$ as the Doppler resolution metric. We present the following resource allocation problem to identify the Pareto optimal point.
\vspace*{-1mm}
\begin{subequations}
\begin{align}
 \underset{m_0,n_0,...,m_K,n_K }
{\text{max}}&
 \frac{\alpha_0 m_0 n_0}{R_{\text{max}}} R_0 -  \sum_{k=1 }^{K}( \frac{\gamma_{k,1} }{d_{\text{max}}~ m_k} \frac{c}{\Delta f} \notag \\
&+ \frac{\gamma_{k,2} }{\nu_{\text{max}}~ n_k} \frac{1}{T_s} ),  \label{st.0}\\
\textrm{s.t. }&m_i \in \{1,...,M_{\text{max}}-1\}, \textrm{  } i = 0,...,K, \\
&  n_i \in \{1,...,N_{\text{max}}-1\},\textrm{  } i= 0,...,K,  \\
& \sum_{i=0}^K m_i = M_{\text{max}}, \\
& \sum_{i=0}^K n_i = N_{\text{max}}, 
\end{align}
\end{subequations}
where $\alpha_0$, $\gamma_{k,1}$, and $\gamma_{k,2}$ represent the communication weight, range resolution weight, and Doppler resolution weight of target $k$, respectively. These values can be selected based on the priorities of the application, ensuring that $\alpha_0 + \sum_{k=1}^K (\gamma_{k,1} +\gamma_{k,2}) = 1$. $M_{\text{max}}$ and $N_{\text{max}}$ are available resources in frequency and time domain, respectively. $R_{\text{max}}$, $d_{\text{max}}$, and $\nu_{\text{max}}$ denote the maximum achievable throughput, range resolution, and Doppler resolution, respectively, when all available resources in the time i.e., $N_{\text{max}}$, and frequency i.e., $M_{\text{max}}$, domains are allocated for communication data transmission or range and Doppler estimation. Additionally, $m_0$ and $n_0$ represent the number of PRBs and slots, respectively, allocated for communication, while $m_k$ and $n_k$ denote the number of PRBs and slots allocated for range and Doppler estimation of the target $k$.
By solving this optimization problem, we can identify the Pareto optimal point for resource allocation between communication and sensing.

\begin{figure}[!t]
    \centering
    \begin{subfigure}{\columnwidth}
        \centering
        \includegraphics[width=\linewidth]{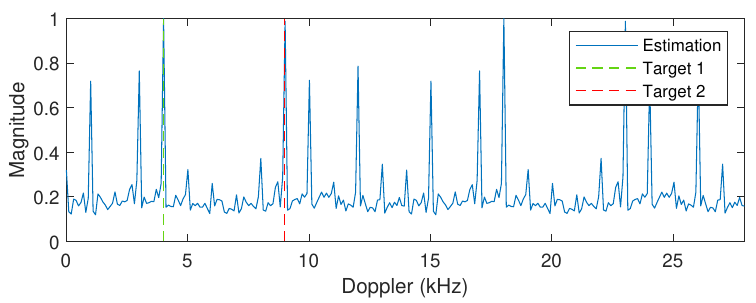}
        \vspace{-0.5cm}
        \caption{Doppler estimation with PRS only.}
         \vspace{0.5cm}
        \label{figa:PRS4}
    \end{subfigure}
    \begin{subfigure}{\columnwidth}
        \centering
        \includegraphics[width=\linewidth]{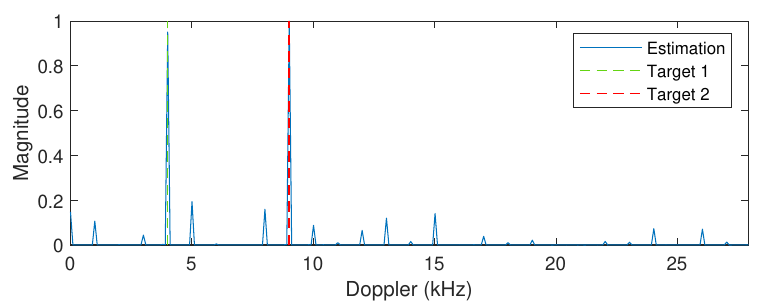}
        \caption{Doppler estimation with PRS and DMRS combined.}
        \label{figb:PRS4_DMRS}
    \end{subfigure}
     \caption{Doppler ambiguity elimination with the presence of 2 targets with $K_{comb} = 4$, $f_c = 4$GHz and $SCS=30$kHz.}
    \label{fig:Doppler_comb4}
\end{figure}

\section{Simulation Results \label{results}}
\begin{figure}[!t]
    \centering
    \begin{subfigure}{\columnwidth}
        \centering
        \includegraphics[width=\linewidth]{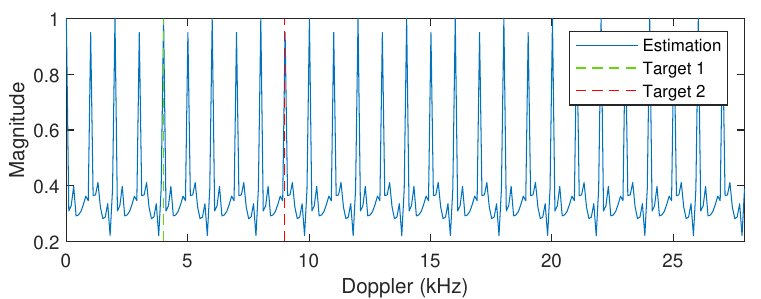}
        \vspace{-0.5cm}
        \caption{Doppler estimation with PRS only.}
         \vspace{0.5cm}
        \label{figa:PRS12}
    \end{subfigure}
    \begin{subfigure}{\columnwidth}
        \centering
        \includegraphics[width=\linewidth]{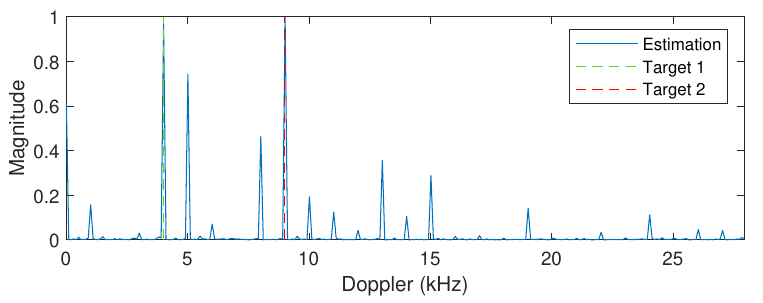}
        \caption{Doppler estimation with PRS and DMRS combined.}
        \label{figb:PRS12_DMRS}
    \end{subfigure}
     \caption{Doppler ambiguity elimination with the presence of 2 targets with $K_{comb} = 12$, $f_c = 4$GHz and $SCS=30$kHz.}
    \label{fig:Doppler_comb12}
\end{figure}

\begin{figure}[!t]
\centering
\mbox{\includegraphics[width=\linewidth]{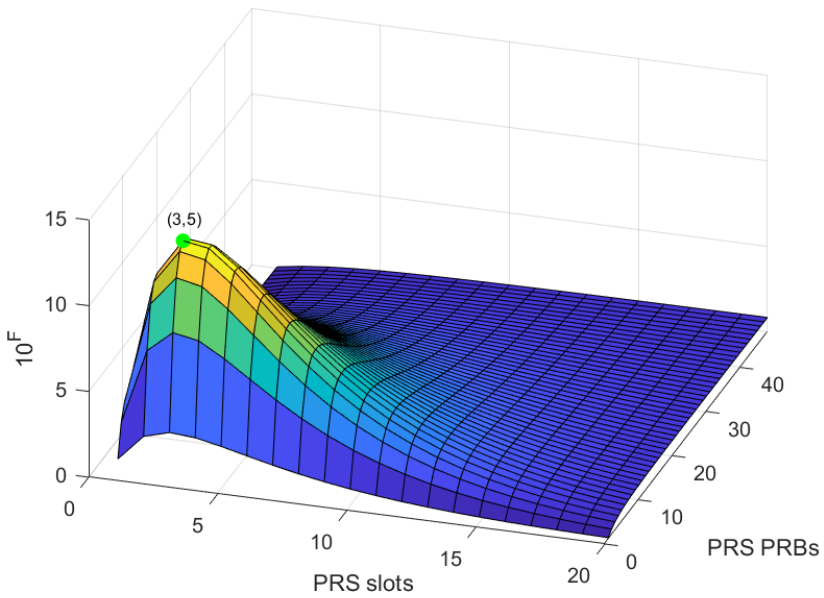}}
\caption{Pareto optimal point in time and frequency resource allocation between communication and sensing.}
\label{pareto}\vspace{-5mm}
\end{figure}

To validate our approach, we exploited MATLAB 5G toolboxes, ensuring compatibility with 3GPP standards. We conducted simulations for a scenario where two targets caused the Doppler frequency shift of $4$kHz and $9$kHz. The central frequency was set to $4$GHz with a subcarrier spacing (SCS) of $30$kHz, and the SNR is $0$dB. PRS number of symbols is $12$, DMRS configuration type is $2$, the number of front-loaded DMRS symbols is $2$ with $1$ additional DMRS symbol position which make $4$ symbols in total, PDSCH mapping type is "A", Code rate used to calculate transport block sizes is $490/1024$ based on \cite{3gpp2018nr1}. Virtual resource block (VRB) bundle size is $4$, VRB to PRB interleaving is disabled, and modulation is 16 quadrature amplitude modulation (16-QAM) for PDSCH and $\beta_1=\beta_2 =1$. In Fig. \ref{figa:PRS4}, we can observe the Doppler ambiguity using PRS only, with the real and false peaks of Doppler frequency shifts. However, in Fig. \ref{figb:PRS4_DMRS}, the magnitudes of the fake Doppler frequencies are significantly suppressed by using the proposed method, and we can easily select the two highest peaks originating from two targets. It is worth mentioning that the number of the targets can be estimated
by model order selection based on information-theoretic criteria \cite{mariani2015model}. In Fig. \ref{figa:PRS12}, we can see an extreme Doppler ambiguity when $K_{comb} = 12$ while
Fig. \ref{figb:PRS12_DMRS} shows the performance of the proposed method with extreme ambiguity in Doppler estimation. In this case, we can still easily select the highest two peaks and estimate the Doppler frequency shifts of the targets correctly. Once the Doppler ambiguity is mitigated, the Pareto optimal point, as illustrated in Fig. \ref{pareto} in time and frequency resource allocation can be found. $F$ in Fig. \ref{pareto} is the objective function of the optimization problem in \eqref{st.0}. In this simulation, we set $M_{\text{max}} = 48$, $N_{\text{max}} = 20$, $K=1$, $\alpha_0 = 2$, and $\gamma_{1,1} = \gamma_{1,2} = 1$. As illustrated in Fig. \ref{pareto}, the Pareto optimal point is attained by allocating $3$ slots and $5$ PRBs for PRS, with the remaining resources allocated to PDSCH.

\section{Conclusion}
In this work, we addressed the Doppler ambiguity appeared using PRS  by exploiting DMRS within PDSCH for unconnected targets in ISAC framework. We proposed a novel method to alleviate Doppler ambiguity, which is fully compliant with 3GPP standards and can be integrated into existing 5G networks. Furthermore, our approach enables sensing without imposing additional overhead, requiring modifications to gNB configurations, or compromising Doppler estimation accuracy, while also enhancing the maximum detectable Doppler shift. Finally, we introduced a resource allocation problem between PDSCH and PRS to determine the Pareto optimal point between communication and sensing while effectively eliminating Doppler ambiguity.
 
\bibliographystyle{IEEEtran}
\bibliography{ref}

\end{document}